\documentclass[aps, prd, superscriptaddress, nofootinbib, preprintnumbers,twocolumn]{revtex4}
\usepackage{amsmath}
\usepackage{graphicx}

\newcommand{\chushi}[1]{}

\begin{document}
 \preprint{MISC-2012-14}
 \title{{\bf Is 125 GeV techni-dilaton found at LHC?}
 \vspace{5mm}}
\author{Shinya Matsuzaki}\thanks{
      {\tt synya@cc.kyoto-su.ac.jp} }
      \affiliation{ Maskawa Institute for Science and Culture, Kyoto Sangyo University, Motoyama, Kamigamo, Kita-Ku, Kyoto 603-8555, Japan.}
\author{{Koichi Yamawaki}} \thanks{
      {\tt yamawaki@kmi.nagoya-u.ac.jp}}
      \affiliation{ Kobayashi-Maskawa Institute for the Origin of Particles and 
the Universe (KMI) \\ 
 Nagoya University, Nagoya 464-8602, Japan.}
\date{\today}

\begin{abstract}
A new particle at around 125 GeV has been observed at the LHC, 
which we show could be identified with the techni-dilaton (TD) 
predicted in the walking technicolor and thus should be an evidence of walking technicolor.  
The TD is a pseudo Nambu-Goldstone boson for the approximate scale symmetry 
spontaneously broken by techni-fermion condensation, with its lightness being ensured by 
the approximate scale invariance of the walking technicolor. 
We test  the goodness-of-fit of  the TD signatures 
using the presently available LHC data set,  and show that the 125 GeV TD is actually favored 
by the current data to explain the reported signal strengths in the  global fit  
as well as in each channel including the coupling properties,  
most notably the somewhat large diphoton event rate. 
\end{abstract} 
\maketitle

\section{Introduction} 

On July 4, the ATLAS and CMS groups~\cite{070412} have reported 
observation of a new boson at around 125 GeV 
in search for the decay channels such as $\gamma\gamma$, $WW^*$ and $ZZ^*$.   
Though the statistical uncertainty is still large, 
the current data on the diphoton event rate~\cite{ATLAS-CONF-2012,CMS-PAS-HIG} exhibit the signal strength 
about two times larger than that expected from the standard model (SM) Higgs boson, which 
may imply the observation of a new boson beyond the SM.

One such a possibility is the techni-dilaton (TD), a composite scalar boson, predicted in 
the walking technicolor (WTC)~\cite{Yamawaki:1985zg,Bando:1986bg} 
with an approximately scale-invariant (conformal) 
 gauge dynamics and  a large anomalous dimension
$\gamma_m =1$ (The WTC was subsequently studied  without notion of anomalous dimension and  scale invariance/TD.~\cite{Akiba:1985rr}).  
The TD  is a pseudo Nambu-Goldstone boson for the spontaneous breaking of 
the approximate scale symmetry triggered by techni-fermion condensation 
and hence its lightness, say 125 GeV,  
would be protected  
by the approximate scale symmetry inherent to the WTC.
Thus the discovery of TD would imply the discovery of the WTC.

In Refs.~\cite{Matsuzaki:2011ie,Matsuzaki:2012gd,Matsuzaki:2012vc}
we studied the LHC signatures of the TD. 
Particularly 
in Ref.~\cite{Matsuzaki:2012vc}
we have shown that the 125 GeV TD 
can be consistent with the currently reported diphoton signal as well as 
other signals such as $WW^*$ and $ZZ^*$, etc..

In this article~\footnote{
Throughout this paper as well as the previous ones~\cite{Matsuzaki:2011ie,Matsuzaki:2012gd,Matsuzaki:2012vc},
 our calculation is based on the (improved) ladder Schwinger-Dyson (SD) equation, which in fact explicitly shows 
the chiral phase transition 
 so-called ``conformal phase transition''~\cite{Miransky:1996pd} 
where the composite scalar does become massless when approaching the critical point from the broken phase, 
though not 
from the unbroken phase (conformal window). 
The ladder-like calculation can only  evaluate a combination (i.e., a product) 
of the mass $M_\phi$ and the decay  constant $F_\phi$ of TD through the  Partially Conserved Dilatation Current  (PCDC) relation, 
Eqs.(\ref{Vac2}),(\ref{Vac}), and  hence such a massless TD limit in the  conformal phase transition is realized  only when $F_\phi/m_F$ diverges (TD  gets decoupled)~\cite{Hashimoto:2010nw}. 
Although there is non-decoupled massless limit in the ladder calculations, we may adjust the TD mass as low as 125 GeV freely to predict the decay constant or the TD coupling by using this ladder-estimated PCDC, the same strategy as done before~\cite{Matsuzaki:2011ie,Matsuzaki:2012gd,Matsuzaki:2012vc}. 
} ,
 we extend the previously reported analysis~\cite{Matsuzaki:2012vc} on the 125 GeV TD 
by testing the goodness-of-fit of  the TD signatures,  based on
the presently available LHC data set~\cite{ATLAS-CONF-2012,CMS-PAS-HIG}. 
We show that the 125 GeV TD is actually favored (slightly better than the SM Higgs) 
by testing the goodness-of-fit of the current data: 
The large diphoton event rate is achieved due to the presence of extra 
techni-fermion loop contributions,  a salient feature of the techni-dilaton noted in the previous papers~\cite{Matsuzaki:2011ie,Matsuzaki:2012gd,Matsuzaki:2012vc}. 
It is also shown that, lacking such extra fermion contributions, a similar dilaton model~\cite{Goldberger:2007zk} and a typical 
radion model~\cite{Randall:1999ee} analyzed recently for LHC~\cite{Coleppa:2011zx,
Barger:2011qn}  tend to be disfavored by the data, mainly due to 
the suppressed diphoton rate, in accord with other similar analyses~\cite{Carmi:2012in}.

\section{TD couplings} 

The TD couplings to techni-fermions and the SM particles have been 
discussed in detail in Ref.~\cite{Matsuzaki:2012vc}, 
which are completely fixed by Ward-Takahashi identities for the dilatation 
current coupled to the TD. 
  Those couplings are collected in an effective nonlinear Lagrangian invariant under 
the scale symmetry~\cite{Matsuzaki:2012vc}: 
\begin{eqnarray} 
{\cal L} &=& 
\frac{F_\pi^2}{4} \chi^2 {\rm Tr}[{\cal D}_\mu U^\dag {\cal D}^\mu U] 
+ \frac{F_\phi^2}{2} (\partial_\mu \chi)^2 
\nonumber \\ 
&& - m_f \left( \left( \frac{\chi}{S} \right)^{2-\gamma_m} \cdot \chi  \right) \bar{f}f 
\nonumber \\ 
&& 
 + \log \left(\frac{\chi}{S} \right) \left\{ 
 \frac{\beta_F(g_s)}{2g_s} G_{\mu\nu}^2 + \frac{\beta_F(e)}{2e} F_{\mu\nu}^2 
\right\}
+ \cdots 
\,, \label{L}
\end{eqnarray} 
where $\chi=e^{\phi/F_\phi}$ with the TD field $\phi$ and decay constant $F_\phi$, 
and $U=e^{2i \pi/F_\pi}$ with the techni-pion fields $\pi$ and decay constant $F_\pi$ 
related to the electroweak (EW) scale $v_{\rm EW} \simeq 246$ GeV by $F_\pi=v_{\rm EW}/\sqrt{N_D}$, 
in which $N_D$ denotes the number of EW doublets formed by the techni-fermions. 
The SM gauge covariant ${\cal D}_\mu U$ term includes the weak gauge boson masses in the usual manner. 
A spurion field  $S$ has been introduced so as to compensate the scale invariance 
explicitly broken by the techni-fermion condensation itself, 
where $S$ is set to 1 after all calculations as in the case of other spurion methods.

The TD Yukawa coupling to the SM $f$-fermion arises from the second line of Eq.(\ref{L}) as~\cite{Bando:1986bg},
\begin{equation} 
 -  \frac{(3-\gamma_m) m_f}{F_\phi} \, \phi \bar{f}f 
\,, 
\end{equation}
along with scale dimension of techni-fermion bilinear operator 
$(3-\gamma_m)$, where the anomalous dimension is $\gamma_m \simeq 1$ in WTC,  which is crucial to
obtain the realistic mass of the SM fermions of the first and the second generations without suffering from the FCNC 
(flavor-changing neutral current) problems~\footnote{
Other issues such as $S$ and $T$  parameters 
may be improved in the WTC~\cite{Appelquist:1991is,Harada:2005ru}, which 
has been implied also in holographic analyses where $S< 0.1$, say,  can easily be arranged~\cite{Haba:2008nz,Haba:2010hu}. 
See also {\it Note added}.
Even if  WTC in isolation cannot overcome this problem, 
the combined dynamical system including the SM fermion mass generation such as the extended TC (ETC)  
dynamics~\cite{Dimopoulos:1979es} may resolve the problem, 
in much the same way as the solution (``ideal fermion delocalization'')~\cite{Cacciapaglia:2004rb} 
in the Higgsless models which simultaneously
adjust the $S$ and $T$ parameters by incorporating the SM fermion mass profile. }.  
However it was known for long time that it is not enough for the mass of the third-generation 
SM $f$-fermions like $t, b, \tau$: 
A simplest resolution would be the strong ETC model~\cite{Miransky:1988gk} 
having much larger anomalous dimension $1<\gamma_m <2$ due to the strong effective four-fermion coupling
from the ETC dynamics 
in addition to the walking gauge coupling.  
As was prescribed in {\it Note added} of Ref.~\cite{Matsuzaki:2012vc},  
here we take $\gamma_m \simeq$ 2, i.e., $(3-\gamma_m) \simeq 1$, 
as in the strong ETC model~\cite{Miransky:1988gk} 
for the third-generation SM $f$-fermions like $t, b, \tau$  
which are relevant to the current LHC data.

 From the Lagrangian Eq.(\ref{L}) one thus easily sees that 
 the TD couplings to $W$ and $Z$ bosons and fermions 
are related to those of the SM Higgs by a simple scaling: 
\begin{eqnarray} 
  \frac{g_{\phi WW/ZZ}}{g_{ h_{\rm SM} WW/ZZ }} 
  &=& \frac{v_{\rm EW}}{F_\phi} 
  \,, \nonumber \\ 
  \frac{g_{\phi ff}}{g_{h_{\rm SM} ff}} 
  &=& \frac{v_{\rm EW}}{F_\phi} 
  \,, 
\qquad {\rm for} \quad f=t,b,\tau 
\,.  \label{scaling}
\end{eqnarray} 

 In addition to the above scaling, 
the couplings to gluon and photon ($G_{\mu\nu}^2$ and $F_{\mu\nu}^2$ terms in Eq.(\ref{L}))  
involve the beta functions, $\beta_F(g_s)$ and $\beta_F(e)$, induced from 
$F$-techni-fermion loops~\cite{Matsuzaki:2011ie}.  This is the most crucial point~\cite{Matsuzaki:2012vc} that distinguishes  the TD from other similar models of dilaton/radion~\cite{Goldberger:2007zk,Randall:1999ee}. 
Here we shall employ the one-family model (1FM) with $N_D=4$ as a WTC setting (see the later discussions) for  
the $SU(N_{\rm TC})$ gauge group~\footnote{
Most of the variants of the WTC have a tendency similar to the one studied here, 
except for the models without colored weak-doublets such as 
the ``one-doublet model'' (with additional singlet techni-fermions),
 which was shown to be invisible at LHC~\cite{Matsuzaki:2011ie,Matsuzaki:2012gd,Matsuzaki:2012vc}. 
 A variant of the one-doublet model, 
``Minimal Walking Technicolor"~\cite{Sannino:2004qp},  a simple WTC to minimize the S parameter
based on the higher TC representation instead of the fundamental one, will also have a light TD, what they called
``composite Higgs". 
Such a ``composite Higgs" would have a quite different LHC phenomenology from  
that given here, 
if the decay constant of the ``composite Higgs" is the same as $F_\pi=v_{\rm EW}$ as assumed in the 
literature~\cite{Sannino:2004qp}, in contrast to our estimate of
$F_\phi$ in Eq.(\ref{vals}) (with $N_{\rm D}=1$)
(See also {\it Note added} in the end of this paper). }. 
 Noting that the scale anomaly-related vertices $\phi$-$\gamma(g)$-$\gamma(g)$ 
 are dominated by the infrared region of the loop integral, 
one can see that the beta functions $\beta_F(g_s)$ and $\beta_F(e)$ 
actually coincide with
those evaluated perturbatively at the one-loop level, similarly to the axial anomaly which is exactly saturated by the
one-loop~\footnote{
 As was discussed in Ref.~\cite{Matsuzaki:2012vc}, 
in the ladder approximation, 
the ultraviolet region of the relevant loop integral $I$ for the $\phi$-$\gamma(g)$-$\gamma(g)$ vertex  
is highly suppressed as 
$ 
I \sim   \int d^4 p \, \frac{\Sigma(p^2)^2}{p^4} \sim \int d^4 p\, p^{2\gamma_m -8}$,  
where $\Sigma(p^2)$ denotes the nonperturbative mass function. 
Hence it is dominated by the infrared region in which the TD-techni-fermion-techni-fermion vertex $\chi_{\phi FF}(p)$
as well as the mass function $\Sigma(p)$ are almost constant to be $\sim m_F/F_\phi$ and $m_F$, respectively. 
This results in the beta function $\beta_F(e/g_s)$ 
in the $\phi$-$\gamma(g)$-$\gamma(g)$ vertex, which coincides 
with that 
calculated at one-loop level of the perturbation with constant mass of the techni-fermion (which was generated nonperturbatively).  
}:
\begin{eqnarray} 
  \beta_F(g_s) &=& \frac{g_s^3}{(4\pi)^2} \frac{4}{3} N_{\rm TC} \,,
\nonumber \\ 
  \beta_F(e) &=& \frac{e^3}{(4\pi)^2} \frac{16}{9} N_{\rm TC} 
\,.   \label{betas}
\end{eqnarray} 
    We thus find the scaling from the SM Higgs for the couplings to 
$gg$ and $\gamma\gamma$, which can approximately be expressed 
 at around 125 GeV as  
\begin{eqnarray} 
\frac{g_{\phi gg}}{g_{h_{\rm SM} gg}} 
&\simeq & 
\frac{v_{\rm EW}}{F_\phi} 
\cdot 
\left( (3-\gamma_m) + 2 N_{\rm TC} \right)   \,,
\nonumber \\ 
\frac{g_{\phi \gamma\gamma}}{g_{h_{\rm SM} \gamma\gamma}} 
&\simeq & 
\frac{v_{\rm EW}}{F_\phi} 
\cdot 
 \left( \frac{63 -  16(3-\gamma_m)}{47} - \frac{32}{47} N_{\rm TC} \right)  
\,,  \label{g-dip-dig}
\end{eqnarray} 
where in estimating the SM contributions  
we have incorporated only the top and $W$ boson loop contributions. Note that  since we used $\gamma_m =2$ 
for the top quark Yukawa coupling as noted before (as well as {\it Note added} of
Ref.~\cite{Matsuzaki:2012vc}),
 Eq.(\ref{g-dip-dig}) is slightly different from  Eq. (51)  of  \cite{Matsuzaki:2012vc} 
(which used $\gamma_m=1$) for the numerical analysis in the present work.   
The full expressions for these couplings are presented in Appendix A of Ref.~\cite{Matsuzaki:2012vc}.

  As seen from Eqs.(\ref{scaling}) and (\ref{g-dip-dig}), 
once the ratio $v_{\rm EW}/F_\phi$ is fixed, 
the TD LHC phenomenological study can be made just by quoting 
the SM Higgs coupling properties. 
 The TD decay constant $F_\phi$ can actually be related to the TD mass $M_\phi$ 
through the 
PCDC -- which is analogous to 
the PCAC (partially conserved axialvector current) 
relation for the QCD pion -- involving the vacuum energy density ${\cal E}_{\rm vac}$: 
\begin{equation} 
  F_\phi^2 M_\phi^2 = - 16{\cal E}_{\rm vac} 
\,. \label{Vac2}
\end{equation} 
 The vacuum energy density ${\cal E}_{\rm vac}$ is saturated by 
 the techni-gluon condensation induced from the techni-fermion condensation, 
 so is generically expressed as 
 \begin{equation} 
   {\cal E}_{\rm vac} 
   = - \kappa_V \left( \frac{N_{\rm TC} N_{\rm TF}}{8\pi^2} \right) m_F^4 
   \,, \label{Vac}
 \end{equation}
where $m_F$ denotes the dynamical techni-fermion mass and $N_{\rm TF}=2 N_{\rm D} + N_{\rm EW-singlet}$ including 
the number of dummy techni-fermions, $N_{\rm  EW-singlet}$, which are singlet under the SM gauges. 
The overall coefficient $\kappa_V$ is determined once a straightforward nonperturbative calculation is made. 
The dynamical techni-fermion mass $m_F$ can, on the other hand, be related to 
the techni-pion decay constant $F_\pi$:  
\begin{equation} 
  F_\pi^2 = \kappa_F^2 \frac{N_{\rm TC}}{4 \pi^2} m_F^2 
\,,   \label{PS}
\end{equation}
with the overall coefficient $\kappa_F$ and the property of $N_{\rm TC}$ scaling taken into account.

The values of $\kappa_V$ and $\kappa_F$ may be quoted from the latest result~\cite{Hashimoto:2010nw} on a 
ladder Schwinger-Dyson analysis for a modern version of 
WTC~\cite{Lane:1991qh,Appelquist:1996dq, Miransky:1996pd}: 
\begin{equation} 
 \kappa_V \simeq 0.7 \,, \qquad 
\kappa_F \simeq 1.4 
\,,  \label{kappas}
\end{equation} 
where $\kappa_F$ has been estimated based on the Pagels-Stokar formula~\cite{Pagels:1979hd}.  
In that case $N_{\rm TF}$ is fixed by the criticality condition 
for the walking regime as~\cite{Appelquist:1996dq} 
\begin{equation} 
 \frac{N_{\rm TF}}{4 N_{\rm TC}} \simeq 1 
 \,.
 \label{criticality}
\end{equation} 
The estimated values in Eqs.(\ref{kappas}) and (\ref{criticality}) 
are based on ladder approximation which are subject to certain uncertainties up to 30\%  
observed for the critical coupling and hadron spectrum in QCD~\cite{Appelquist:1988yc}.   
 We may include this 30\% uncertainty in estimation of each independent factor $\kappa_V$, $\kappa_F^2$   
and the criticality condition $N_{\rm TF}/(4N_{\rm TC})$. 
  Putting these all together, we thus estimate $v_{\rm EW}/F_\phi$ as  
\begin{eqnarray} 
\frac{v_{\rm EW}}{F_\phi}
&\simeq& 
(0.1 - 0.3) \times  \left( \frac{N_D}{4} \right) \left( \frac{M_\phi}{125\,{\rm GeV}} \right) 
\,, \label{vals}
\end{eqnarray}
and $m_F \simeq (320 - 420){\rm GeV} \sqrt{3/N_{\rm TC}}$ 
for the 1FM with $F_\pi \simeq 123$ GeV.

 From Eqs.(\ref{scaling}) and (\ref{vals}) we see that 
{\it the TD couplings to $WW, ZZ$ and $f\bar{f}$ 
are substantially smaller than those of the SM Higgs. 
 On the other hand, the TD couplings to $gg$ and $\gamma\gamma$ in Eq.(\ref{g-dip-dig}) have 
 extra factors $(1 + 2 N_{\rm TC})$ and $(1 - 32 N_{\rm TC}/47)$ 
coming from techni-fermions as well as the $W$ and top quarks 
carrying the QCD color and electromagnetic charges}. 
  The gluon fusion (GF) production at the LHC thus becomes larger than the SM Higgs case 
  due to this extra factor, while other productions such as vector boson fusion (VBF) 
and vector boson associate (VBA) productions are significantly suppressed.  
Note also that the coupling to $\gamma\gamma$ becomes enhanced when 
 $N_{\rm TC} \ge 4$ where the total fermionic loop contributions get large enough to 
 overcome the destructive $W$-loop contributions as well as to compensate the overall suppression by $v_{\rm EW}/F_\phi$.

\section{TD signal at LHC}

As done in Refs.~\cite{Matsuzaki:2012gd,Matsuzaki:2012vc}, 
we can calculate   
the TD production cross sections and decay widths including loop contributions from the 
SM particles, by quoting the corresponding formulas for the SM Higgs~\cite{Spira:1997dg}. 
(The explicit expressions of the formulas for TD are given in Ref.~\cite{Matsuzaki:2012vc}.) 
Here we focus on the GF and VBF productions for decay channels to $WW^*$, $ZZ^*$, $\tau^+\tau^-$ and $\gamma\gamma$, 
and VBA production for $b\bar{b}$ channel, to which 
  the ATLAS and CMS experiments have so far reported the 
significant data in search for the SM Higgs~\cite{ATLAS-CONF-2012,CMS-PAS-HIG}.

  The signal strength $\mu \equiv \sigma/\sigma_{\rm SM}$ for $b\bar{b}$ channel 
 at $\sqrt{s}=7, 8$ TeV  is thus evaluated as     
\begin{eqnarray} 
    \mu_{b\bar{b}}  
&=& 
\frac{\sigma_{\rm VBA}^{\phi}(s)}
{\sigma_{\rm VBA}^{h_{\rm SM}}(s)}  
\frac{BR(\phi \to b\bar{b})}{BR(h_{\rm SM} \to b\bar{b})}  
\nonumber \\ 
&=& 
\frac{\sigma_{W \phi}(s) + \sigma_{Z \phi}(s)}{\sigma_{W h_{\rm SM}}(s) + \sigma_{Z h_{\rm SM}}(s) }  
\frac{BR(\phi \to b\bar{b})}{BR(h_{\rm SM} \to b\bar{b})}  
\,. 
\end{eqnarray}
 For $X=WW^*$, $ZZ^*$ and $\tau^+\tau^-$ channels we take the signal strengths 
to be inclusive:     
\begin{equation} 
  \mu_{X} = 
\frac{\sigma_{\rm GF}^{\phi}(s) + \sigma_{\rm VBF}^{\phi}(s)}
{\sigma_{\rm GF}^{h_{\rm SM}}(s) + \sigma_{\rm VBF}^{h_{\rm SM}}(s)}  
\frac{BR(\phi \to X)}{BR(h_{\rm SM} \to X)}  
\,.  
\end{equation}
On the other hand, 
$\gamma\gamma + 0j$  and $\gamma\gamma + 2j$ channels are treated to be exclusive by distinguishing 
the TD production processes: 
\begin{eqnarray} 
  \mu_{\gamma\gamma 0j} 
&=& 
\frac{\sigma_{\rm GF}^{\phi}(s)}
{\sigma_{\rm GF}^{h_{\rm SM}}(s)}  
\frac{BR(\phi \to X)}{BR(h_{\rm SM} \to X)}  
\,, \nonumber \\ 
  \mu_{\gamma\gamma 2j} 
&=&  
\frac{\xi_{\rm GF} \cdot \sigma_{\rm GF}^{\phi}(s) + \xi_{\rm VBF} \cdot \sigma_{\rm VBF}^{\phi}(s)}
{\xi_{\rm GF} \cdot \sigma_{\rm GF}^{h_{\rm SM}}(s) + \xi_{\rm VBF} \cdot \sigma_{\rm VBF}^{h_{\rm SM}}(s)}  
\nonumber \\  
&& 
\times 
\frac{BR(\phi \to \gamma\gamma)}{BR(h_{\rm SM} \to \gamma\gamma)}  
\,,
\end{eqnarray}
where the corresponding acceptances multiplied by 
dijet tag efficiencies $\xi_{\rm GF}$ and $\xi_{\rm VBF}$ are read off from Refs.~\cite{ATLAS-CONF-2012,CMS-PAS-HIG}.  
We then combine the 7 TeV and 8 TeV signal strengths with the luminosities accumulated 
for each event category. 
It turns out that $\mu_{\gamma\gamma 0j}$  
can be enhanced by the factors both from the $gg$ and $\gamma\gamma$ couplings (See Eq.(\ref{g-dip-dig})), 
which can compensate the smallness of $(v_{\rm EW}/F_\phi)$ in Eq.(\ref{vals}) 
for a moderately large $N_{\rm TC}$~\cite{Matsuzaki:2012vc}.

{\it This feature is in sharp contrast to other similar dilaton models such as EW pseudo-dilaton~\cite{Goldberger:2007zk} and 
Randall-Sundrum (RS) radion~\cite{Randall:1999ee}}, 
where extra contributions beyond the SM such as techni-fermions 
are absent so that their diphoton rates are not enhanced, to be disfavored by the current diphoton data, 
in accord with other similar analyses in Ref.~\cite{Carmi:2012in}, as 
will be shown more explicitly below.

We shall test the goodness-of-fit of the TD, based on the $\chi^2$ function: 
\begin{equation} 
  \chi^2 = \sum_{i \in {\rm events}} \left( \frac{\mu_i - \mu_i^{\rm exp} }{\sigma_i} \right)^2  
  \,, 
\end{equation}
where $\mu_i^{\rm exp}$ denote the best-fit strengths for each channel reported in Refs~\cite{ATLAS-CONF-2012,CMS-PAS-HIG} 
and $\sigma_i$ the corresponding one sigma errors. 
Taking $(v_{\rm EW}/F_\phi)$ as a free parameter so as to satisfy the theoretically expected range in Eq.(\ref{vals}), 
 in Fig.~\ref{TD125-fit} we plot the $\chi^2$ function for the 125 GeV TD in 
the 1FM with $N_{\rm TC}=4,5$. 
 The best-fit values are as follows: 
\begin{eqnarray} 
\begin{array}{c|cc} 
\hline 
 N_{\rm TC}  & \hspace{15pt}  (v_{\rm EW}/F_\phi)_{\rm best} \hspace{15pt}  
& \hspace{15pt}  \chi^2_{\rm min}/{\rm d.o.f}  \hspace{15pt}  \\    
\hline 
4 & 0.22 & 12/13 \simeq 0.9  \\ 
5 & 0.17 & 10/13 \simeq 0.7  \\ 
\hline 
\end{array}
\,, 
\end{eqnarray}
which are compared with the SM Higgs case, $\chi^2_{\rm min}|_{\rm SM} \simeq  14/14 = 1.0$, 
implying that the TD is more favorable than the SM Higgs $(\mu_i=1)$. 
 This nice goodness of fit is due to the significant enhancement in the diphoton channel 
coming from the sector beyond the SM (techni-fermions), 
as was also noted in Ref.~\cite{Ellis:2012hz}: The techni-fermion loop contributions as in Eq.(\ref{g-dip-dig}) 
become large enough to compensate  the smallness of the overall $(v_{\rm EW}/F_\phi)$ in Eq.(\ref{vals}).

 In Fig.~\ref{TD125-fit}  also has been shown a comparison with other similar models,
 EW pseudo-dilaton~\cite{Goldberger:2007zk} 
and RS radion~~\cite{Randall:1999ee}.  
These dilaton/radion scenarios are actually disfavored  mainly 
due to the absence of enhancement of diphoton rate, in sharp contrast to the TD~\footnote{
 A variant of 
 the RS radion such as the one discussed in Ref.~\cite{Cheung:2011nv} could actually be as favorable as the TD 
 because of an accidental enhancement of the diphoton event rate  
induced solely from a choice of the SM sector contributions.}.  
This result is in accord with other similar analyses in Ref.~\cite{Carmi:2012in} for the 
EW pseudo-dilaton and in Ref.~\cite{Giardino:2012dp} for the RS radion performed in light of the 125 GeV LHC events. 
although we have not included the Tevatron results on the $\bar b b$ 
channel~\footnote{
 Even including the Tevatron results for $b \bar{b}$, $WW^*$ and $\gamma\gamma$ channels, 
  the goodness-of-fit of TD does not substantially change so much 
($\chi^2_{\rm min}/{\rm d.o.f} = 19/16 \simeq 1.2$ compared with the SM Higgs case, $\chi^2_{\rm min}/{\rm d.o.f} =18/17 \simeq 1.1$).     
}.

Finally, in Fig.~\ref{mu-fit2} we explicitly compare the best-fit signal strengths 
of TD with those estimated 
by the ATLAS and CMS analyses for each channel.  
Note first that the most precise measurement has currently been done in the diphoton 
channel with 0 jet ($\gamma\gamma 0j$) 
which exhibits about 2 times larger signal strengths than the SM Higgs prediction 
($\mu_{\gamma\gamma 0j} = 1.9 \pm 0.5$ for ATLAS 7TeV + 8TeV and 
$\mu_{\gamma\gamma 0j}=1.7 \pm 0.5$ for CMS 7TeV + 8TeV~\cite{ATLAS-CONF-2012,CMS-PAS-HIG}). 
The $\chi^2$ fit is therefore fairly sensitive to the $\gamma\gamma 0j$ category, 
and hence currently the TD can be more favorable than the SM Higgs due to the 
enhancement of the diphoton rate which happens when $N_{\rm TC} \ge 4$.   
 On the other hand, 
the TD signal strength in the diphoton plus dijet channel ($\gamma\gamma 2j$) 
tends to be smaller than the SM Higgs prediction, simply because of the suppression of the 
overall TD coupling compared to the SM Higgs $(v_{\rm EW}/F_\phi)$ in Eq.(\ref{vals}). 
Similar suppressions are also seen in other exclusive channels like 
 $2l2\nu + 2j$ and $\tau^+\tau^- + 2j$ as well as $b\bar{b}$ originated from the VBA and VBF productions. 
 Thus more precise measurements in such other exclusive events  
would draw a more definite conclusion that the TD is favored or not.

\section{Summary}

In summary, we have tested the goodness-of-fit of 
 the 125 GeV TD signatures, 
using the presently available LHC data set by focusing on 
the decay channels to $b\bar{b}$, $WW^*$, $ZZ^*$, $\tau^+ \tau^-$ and $\gamma\gamma$. 
We found that the 125 GeV TD is actually favored by the current data to explain 
the reported signal strengths including the coupling properties 
involving somewhat large diphoton event rate.

The issue to be studied in the future is a reliable nonperturbative  estimate of the TD mass:  There has been only a crude
estimate based on the ladder approximation (see references cited in e.g., Ref.\cite{Matsuzaki:2012vc}). The holographic 
method~\cite{Haba:2010hu,Kutasov:2012uq,Matsuzaki:2012xx} may give some hint and the lattice simulations eventually give us the definite answer.

\section*{Acknowledgments}

This work was supported by 
the JSPS Grant-in-Aid for Scientific Research (S) \#22224003 and (C) \#23540300 (K.Y.).  
\vspace*{20pt}

{\it Note added}--- 
After submitting this paper, we further posted a paper~\cite{Matsuzaki:2012xx} (referred to in the original version as ``in preparation")
 to show  
that there exists an exactly massless limit of TD in a holographic model which is a deformation of 
the model successful for the ordinary QCD with $\gamma_m\simeq 0$ to the 
walking case with $\gamma_m=1$. 
Hence the light TD with 125 GeV is naturally realized.  
In contrast to the ladder approximation employed as in the present paper, 
the holographic model in Ref.~\cite{Matsuzaki:2012xx} 
fully incorporates the techni-gluonic dynamical contributions parametrized by the holographic parameter $G$ corresponding to the techni-gluon condensate,
which makes the massless limit be realized at $G \rightarrow \infty$.  
 The light TD mass $\simeq 125$ GeV 
is actually realized by a large gluonic contribution $G \simeq 10$, which 
is compared with the real-life QCD case with $G \simeq 0.25$~\cite{Haba:2010hu,Matsuzaki:2012xx}. 
In such a light TD case the TD decay constant $F_\phi$ turns out to be free from the holographic parameters, 
$F_\phi/F_\pi \simeq \sqrt{2 N_{\rm TF}}$, which implies 
$v_{\rm EW}/F_\phi \simeq 0.2-0.4$ including a typical size of $1/N_{\rm TC}$ corrections~\cite{Matsuzaki:2012xx}. 
This coincides with the ladder estimate in Eq.(\ref{vals}). 
This may imply that in spite of the qualitative difference in the sense that the holography has a massless TD limit while the ladder does not, 
both models 
are reflecting some reality through similar dynamical effects for the particular mass region of the 125 GeV TD. 
 Note also that this holographic model naturally has a small $S$ parameter, say $S<0.1$,  for the one-family model.

\newpage

\begin{widetext}

 \begin{figure}[h] 

\begin{center} 
\includegraphics[width=8cm]{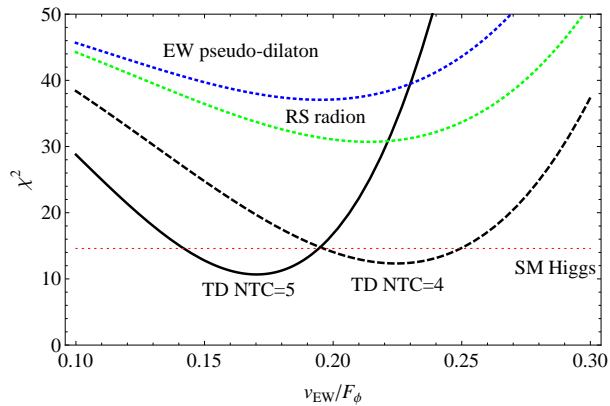} 
\end{center}
\caption{ 
The plot of $\chi^2$ as a function of $v_{\rm EW}/F_\phi$ 
in the case of the 1FM with $N_{\rm TC}=4$ (black dashed curve) 
and $N_{\rm TC}=5$ (black solid curve). 
Comparison with the EW pseudo-dilaton~\cite{Goldberger:2007zk} (blue dotted curve) and RS radion~\cite{Barger:2011qn} 
(green dotted curve) is also shown, along with the SM Higgs case (red dotted line).     
 Here $\chi^2_{\rm min}/{\rm d.o.f}=12/13(10/13)\simeq 0.9(0.7)$ for the TD with $N_{\rm TC}=4(5)$, 
 $\chi^2_{\rm min}/{\rm d.o.f}=37/13 \simeq 2.8$ for the EW pseudo-dilaton and  
 $\chi^2_{\rm min}/{\rm d.o.f}=30/13 \simeq 2.3$ for the RS radion. 
}  
\label{TD125-fit}
\end{figure}

\begin{figure}[h] 

\begin{center} 
\includegraphics[width=10cm]{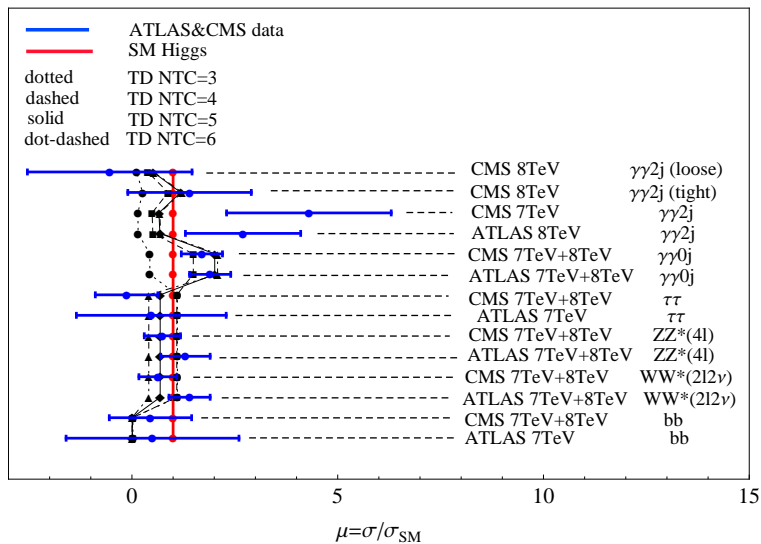} 
\end{center}
\caption{ 
 The best-fit signal strengths of the 125 GeV TD, 
for the decay channels categorized as
$WW^*(2l2\nu)$, $ZZ^*(4l)$, $\tau^+\tau^-$, $\gamma\gamma 0j$ and $\gamma\gamma 2j$~\cite{ATLAS-CONF-2012,CMS-PAS-HIG}. 
}  
\label{mu-fit2}
\end{figure}

\end{widetext}

\end{document}